\documentclass{aa}

\usepackage[]{epsfig}

\def\ltsima{$\; \buildrel < \over \sim \;$}
\def\simlt{\lower.5ex\hbox{\ltsima}}
\def\gtsima{$\; \buildrel > \over \sim \;$}
\def\simgt{\lower.5ex\hbox{\gtsima}}
\def\cgs{{erg cm$^{-2}$ s$^{-1}$}}
\def\ergs{{erg s$^{-1}$}}
\def\cm2{{cm$^{-2}$}}
\def\xdof{{$\chi^{2}$(dof)}}

\def\xred{{$\chi^{2}_{\rm red}$}}

\def\ghard{{$\Gamma_{\rm 2-10}$}}
\def\fhx{{$F_{\rm 2-10}$}}
\def\lum{{$L_{\rm 2-10}$}}
\def\lums{{$L_{\rm 0.5-2}$}}
\def\p1{{Paper I}}
\def\fsx{{$F_{\rm 0.5-2}$}}
\def\xmm{{\em XMM--Newton}}

\def\nh{{N$_{\rm H}$}}
\def\nhz{{N$_{\rm H}^{\rm z}$}}

\def\xmm{{\em XMM--Newton}}

\def\nh{{N$_{\rm H}$}}  
\def\epic{{\em EPIC}}
\def\mosuno{{\em MOS1}}
\def\mosdue{{\em MOS2}}
\def\pn{{\em PN}}
\def\mos{{\em MOS}}

\def\f14{{10$^{-14}$}}
\def\f13{{10$^{-13}$}}
 
\def\rbs{{RBS~315}}
\begin{document}
 
\title{XMM--Newton discovery of soft X--ray absorption\\ in the high-z 
superluminous Blazar RBS~315\thanks{Based on observations
with  XMM-Newton, an ESA Science Mission with instruments and
contributions  directly funded by ESA Member states and the USA
(NASA)}}
 
\author{E.~Piconcelli and M.~Guainazzi}

\titlerunning{XMM-Newton observation of the Blazar RBS~315} \authorrunning{E.~Piconcelli \& M.~Guainazzi}
 
\offprints{Enrico.Piconcelli@sciops.esa.int}
 
\institute{European Space Astronomy Center (ESA), Apartado 50727, E--28080 Madrid, Spain}
 
\date{}
 
\abstract{We present the analysis and the results of a 20 ks
\xmm~observation of the extremely X-ray loud (L$_{\rm X} \approx$ 5
$\times$ 10$^{47}$ \ergs) flat--spectrum radio quasar RBS~315 at a
redshift of 2.69.  This \epic~observation has allowed us to strongly
constrain  the slope of the continuum ($\Gamma$  = 1.23$\pm$0.01) as
well as to discover  the presence of a sharp drop below $\approx$ 2
keV in its spectrum. 
Such a flat photon index and the huge luminosity suggest that the
X-ray emission is due to the low energy tail of the Comptonized
spectrum, produced from plasma in a relativistic jet oriented close
to our line of sight. Even though the hypothesis of a break in the
continuum cannot be completely discarded as an explanation of the soft
X-ray cutoff, the presence of intrinsic absorption appears more plausible. 
 Spectral fits with cold (\nhz~=
1.62$^{+0.09}_{-0.09}$ $\times$ 10$^{22}$ \cm2) and lukewarm (\nhz~= 2.2$^{+0.9}_{-0.3}$
$\times$ 10$^{22}$ \cm2; $\xi$ = 15$^{+38}_{-12}$ \cgs) absorbers are
statistically indistinguishable. Remarkably, our results are very similar to
those reported so far for other absorbed high-$z$ Blazars observed by
\xmm. The existence of this ``homogeneous'' class of jet-dominated
superluminous obscured QSOs at high $z$ therefore could be 
important in the context of the formation and cosmological evolution
of radio-loud objects.   }
 
\maketitle
 
\section{Introduction}

According to Active Galactic Nuclei (AGNs) Unified models (Urry \&
Padovani 1995) a relativistic jet  emitting a beamed
non-thermal continuum is present in radio-loud quasars (QSOs): if the
jet points along our line of view, the QSO is seen as a Blazar.  Due
to strong amplification and collimation in the observer's frame
Blazars appear as the most luminous objects in the sky from radio
through gamma-rays.  As pointed out by Fossati et al. (1998)  the
X-ray spectrum of Blazars hardens with increasing luminosity:  in fact
flat--spectrum radio-loud QSOs (FSRQs), which represent the most
luminous class of Blazars, typically show photon indices $\Gamma
\approx$ 1.3--1.5 in the 2--10 keV band (e.g. Boller et al. 2000;
Donato et al. 2005; Ferrero \& Brinkmann 2003, F03 hereafter)
i.e. flatter than  usually observed in ``normal'' radio-loud
($\langle$\ghard$\rangle \approx$ 1.75; Sambruna et al. 1999) or
radio-quiet QSOs ($\langle$\ghard$\rangle \approx$ 1.9; Piconcelli et
al. 2005a).

The discovery of a low-energy cutoff due to heavy cold absorption
($\sim$ 10$^{22-23}$ \cm2) in the X-ray spectra of some  luminous
high-$z$ FSRQs (Elvis et al. 1994; Cappi et al. 1997; Boller et al. 2000; Yuan et
al. 2000) was therefore unexpected since our line  of sight should not
intercept the obscuring torus. An explanation in terms of obscuration
associated with intervening systems was promptly ruled out by
O'Flaherty \& Jakobsen (1997) and Fiore et al. (1998) on the basis of
statistical arguments,  supporting therefore the hypothesis that the
absorber is physically associated to the QSO.  However, the properties
of the absorbing matter are still poorly known and the observational constraints about its
nature are very sparse.  Unfortunately  the low sensitivity and
limited soft X-rays bandpass of {\it ASCA} and {\it BeppoSAX} data as
well  as the usual high $z$ of the ``obscured'' FSRQs  prevented
deeper investigations about the origin of the absorption features.
Thanks to its unprecedented large collecting area, \xmm~has allowed
the investigations in this field to be re--opened.  Worsley et
al. (2004a, W04 hereafter) claimed for  a warm absorber in the
spectrum  of the Blazar PMN J0525-3343, while F03 found that both a
warm and a cold absorber can fit the data of the Blazar PKS~2126-158.
On the other hand, some recent works based on \xmm~observations (F03;
Grupe et al. 2004; Piconcelli et al. 2005b) do not confirm earlier
{\it ASCA} measurement of a strong absorption toward three high--$z$
QSOs.  

In this Letter we present the analysis of the \xmm~observation of
\rbs~that has allowed  to explore in detail the 0.3--12 keV spectrum
of this source for the first time.  \rbs~was discovered as a powerful
flat-spectrum radio source  ($S_{\rm 6cm}$ = 595 mJy; $\alpha_{\rm
6-20cm}$ = 0.2) by Lawrence et al. (1983) and subsequently identified
as a QSO at $z$ = 2.69 with a radio-loudness parameter 
$R$=$S_{\rm 6cm}/F_{\rm B}$ $\approx$ 6500
in the {\it ROSAT} Bright Survey (RBS; Schwope et al. 2000).  \rbs~was
detected in X-rays by {\it ROSAT} and  {\it RXTE} during the All-Sky
Slew Survey (XSS; Sazonov et al. 2004).  Both X-ray measurements imply
a huge luminosity of $L_X$ \simgt~10$^{47}$ \ergs~and a very flat
hardness ratio, which is a clear indication of a convex spectrum.
\section{Observation and data reduction}
The \xmm~observation of \rbs~took place on 2003 July 25 and  it was
performed with the \epic~\pn~(Struder et al. 2001) and \mos~(Turner et
al. 2001) cameras operating in full-frame mode.  Data were reduced
with $SAS$ 6.1 (Loiseau et al. 2004), using the most updated
calibration files available as of March 1, 2005.  X--ray events
corresponding to patterns 0--12(0--4) for the \mos(\pn)~cameras were
selected.  The event lists were  filtered to ignore periods of high
background flaring  according to the method presented in Piconcelli et
al. (2004) based on the cumulative distribution function of background
lightcurve count-rates.  After this data cleaning, we obtained as net
exposure time 18.3, 21.5, 21.7 ks for \pn,  \mosuno~and \mosdue,
respectively.  
The centroid position of \rbs~in the \epic~image is RA=02$^{h}$ 25$^{m}$ 04$^{s}$.5, 
Dec=$+$18$^{\circ}$ 46$^{\prime}$ 51$^{\prime\prime}$, 3.6$^{\prime\prime}$ away from the 
position of its radio counterpart (Beasley et al. 2002), 
This shifting is, however, smaller than the astrometric error box radius 
of \xmm~($\approx$ 6$^{\prime\prime}$).
Source photons were extracted using a circular region
of radius 36 arcsec centered  at the source position while backgrounds 
were estimated from source--free similar regions close to the source
on the same CCD.  No exposure is affected by pile--up.  Response
matrices and ancillary response files were generated using the RMFGEN
and ARFGEN tools in the $SAS$ software.  Since difference between the
{\em MOS1} and {\em MOS2} response matrices are  a few percent,  we
created a combined \mos~spectrum and response matrix.  The \pn~and
\mos~spectra were then fitted simultaneously.  Both spectra were
grouped to give a minimum of 30 counts per bin in order to apply
$\chi^{2}$ statistics. Given the current calibration uncertainties and
the detector sensitivities (Kirsch et al. 2004), events outside the
0.3--12 keV range were ignored in the \pn~spectrum, while, for the
\mos, we retained the 0.8--10 keV band. The quoted errors on the model
parameters correspond to a 90\% confidence level for one interesting
parameter ($\Delta\chi^2$ = 2.71; Avni 1976). All luminosities are
calculated assuming a $\Lambda$CDM cosmology with ($\Omega_{\rm
M}$,$\Omega_{\rm \Lambda}$) = (0.3,0.7) and a Hubble constant of 70 km
s$^{-1}$ Mpc$^{-1}$.

\section{Results}

\begin{figure}\centering
\epsfig{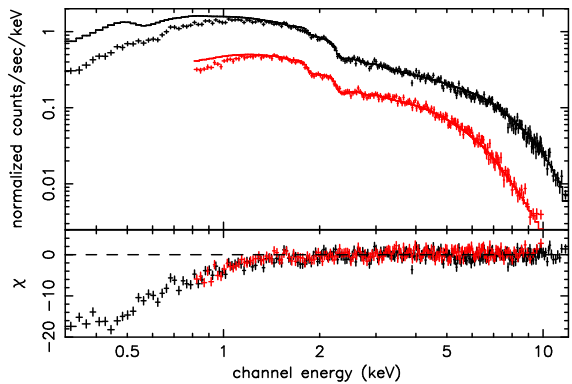}
  \caption{Continuum power-law fit to the hard band of the \pn~(top) and \mos~(bottom) spectra of \rbs~extrapolated over the 0.3--12(0.8--10) keV band for 
\pn(\mos). The lower panel shows the deviations of the observed data from the model in unit of standard deviations.}
   \label{fig1} 
\end{figure}

All the spectral fits were performed with the XSPEC package (version
11.3.0).  The models presented in this Section include an
absorption component due to the line--of--sight Galactic column of
\nh~= 10$^{21}$ \cm2~(Dickey \& Lockman 1990).  We initially fitted
the hard (2--10 keV, corresponding to 7.4--37 keV in the frame of
\rbs) portion of the spectrum with a power law. This fit turned out to
be very good yielding an associated \xred~=1.07 with a resulting
best-fit value of the photon index $\Gamma$=1.24$^{+0.01}_{-0.02}$.
Visual inspection of the data--to--model ratio residuals did not
suggest the presence of any emission/absorption features. However, the
extrapolation of the power law to energies lower than 2 keV clearly
revealed the presence of a deep deficit in the soft range of the
spectrum (Fig.~\ref{fig1}). Fitting the data over the 0.3--12 keV band
with a simple power law (Model PL) yielded a \xdof~1821(455).  We
accounted for the soft X-ray spectral drop by means of a neutral (``cold'')
absorption component in the fitting model.  This fit (model APL
hereafter) resulted in a dramatic statistical improvement i.e. at the
$>$99.99\% confidence level according to an $F-$test once compared
with the model PL ($\Delta\chi^2$ = 1342), with a \xdof~= 479(454).
This fit therefore provides an excellent parametrization of the 
X--ray spectrum of this Blazar with a best--fit value of $\Gamma$ = 1.23$\pm$0.01
and \nhz~=1.62$^{+0.09}_{-0.09}$ $\times$ 10$^{22}$ \cm2~for the spectral index 
and the intrinsic column density 
of the 'cold' absorber, respectively.
Using Model APL we measured a flux of \fsx~= 3.1 $\times$ 10$^{-12}$
\cgs~and \fhx~=  1.6 $\times$ 10$^{-11}$ \cgs~ in the low (0.5--2 keV)
and high (2--10 keV) energy bands, respectively.  After correction for
both Galactic and intrinsic absorption column densities, these
correspond to luminosities of \lums~= 7.5  $\times$ 10$^{46}$
\ergs~and \lum~= 3.5 $\times$ 10$^{47}$ \ergs, respectively.
It is worth to note that if the redshift of the absorber if fixed to $z$=0,
the resulting column density value decreases to \nh($z$=0) = 9.4$\pm$0.1 $\times$ 
10$^{20}$ \cm2. However this fit, with an associated \xdof~= 508(454), is statistically worse 
than Model APL  (i.e. $\Delta\chi^2$ = 29 for the same number of d.o.f.) 
and some residuals, as ratio between model and data, are clearly present in the 
soft X-ray band. The hypothesis that the absorber could be associated with our Galaxy
appears, therefore, quite unlikely.

As W04 suggested the presence of a warm absorber as the most
plausible explanantion of the soft X-ray cutoff in the \xmm~spectrum of
the Blazar PMN~J0525-3343 at $z$ = 4.4, 
we then investigated if it is the case also for RBS~315.
This fit, performed using the {\tt ABSORI} model in XSPEC, is statistically equivalent to that with the APL model, i.e. associated \xdof~= 478(453). The resulting column density of the warm gas was
\nh~= 2.2$^{+0.9}_{-0.3}$ $\times$ 10$^{22}$ \cm2, while the best-fit value ionization parameter
was $\xi$ = 15$^{+38}_{-12}$ \cgs. This value is comparable to what reported by W04.\\ 
An alternative explanation for the convex spectrum in Blazar is the
presence of an underlying break in the continuum. 
 Fossati et al. (1998) have shown that
the spectral energy distribution of Blazars is ``double-humped'' with
two main components due to the synchrotron emission and Inverse
Compton (IC) scattering of relativistic electrons off synchrotron or
ambient ``seed'' photons, respectively. 
In the case of the FSRQs, such as \rbs, the IC hump peaks at $\sim$10-100 MeV: this means that
the spectrum in the \xmm~bandpass is dominated by the Comptonized ``boosted'' emission from the jet.
A spectral break
can originate through  a low-energy  cutoff in the energy distribution
of the radiating electrons or by a sharply peaked external ``seed'' photon
distribution (e.g. Fabian et al. 2001). We then tried an alternative fit modeling the spectrum
with a broken power law.  This parameterization yielded unacceptable fit with an associate \xdof~= 544(453),
 i.e. significantly worse than that derived with model APL.
A model with two (unabsorbed) power laws also turned out in a very poor fit (\xred~\simgt~3). 
\begin{figure}\centering
\epsfig{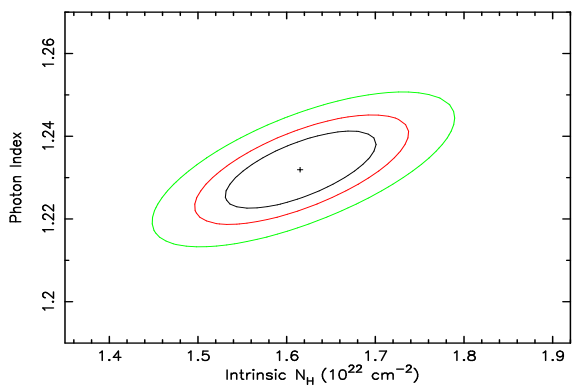}
  \caption{Confidence contour plot showing the photon index against the
rest--frame column density (\nhz) of the absorber. The contours are at 68\%, 90\% and 99\% confidence levels, respectively, for two interesting parameters.}
   \label{fig3} 
\end{figure}
\section{Discussion}

The \epic~observation presented here provides the first good--quality
X-ray spectrum of \rbs. It has allowed us to strongly constrain  the
slope of the continuum ($\Gamma$  = 1.23$\pm$0.01) as well as to
discover  the presence of a sharp drop in the observed spectrum below
1.5 keV.  Such a value of photon index is consistent with previous
X-ray observations of luminous FSRQs ($\langle\Gamma_{\rm 0.3-4.5
keV}\rangle$ = 1.48, Sambruna et al. (1997); see also W04 and F03) and
is also in agreement with the trend found by Fossati et al. (1998) for
a large sample of Blazars according to which the more X-ray luminous
the object is, the flatter X-ray spectrum it possesses.   The explanation
for this behavior is that the X-ray continuum in FSRQs is believed to
be  the low energy tail of the Comptonized spectrum peaking in the
MeV--GeV range and as the peak moves to lower energies with increasing
luminosity, the X-ray spectrum therefore becomes harder.\\

The most important results of our analysis is the discovery of a soft
X-ray spectral  flattening in RBS~315.  In the previous Section we
tested three different models in order to account for this spectral
feature: a power law modified by either a (i) cold  or a (ii) warm
absorption component and (iii) a broken power law.  The latter
provided the worst description of the \xmm~spectrum with an associated
\xred~= 1.2.   Furthermore, as pointed out by Worsley et al. (2004b),
an explanation of the observed cutoff in terms of a spectral break is
not particularly compelling  since such a feature is not observed so
far in any of the nearby FSRQs, for which high-quality X-ray data have
been collected.  An additional problem that further weakens this
scenario is the sharpness of the observed break  as it takes place
over a energy range of a few keVs in the QSO frame.  It would
therefore require an unlikely,  nearly monochromatic ``seed'' photon
distribution (e.g. Ghisellini 1996 for details) to be generated.

Hence, even though the hypothesis of a break in the continuum cannot
be completely discarded, it appears quite unlikely once compared with
the possible presence of absorption in this FSRQ.

On the other hand, the spectral fittings with cold and warm absorber
are statistically indistinguishable: using the present X-ray data  it
is not possible to constrain the ionization state of the absorbing matter better than
$\xi$  = 15$^{+38}_{-12}$ \cgs.
Moreover, the most prominent hallmarks of a warm absorber, typically
the Fe M-shell UTA and the O/Ne absorption edges, are located in the
rest-frame energy interval 0.6--0.9 keV  which is, unfortunately, outside the
observed (i.e. 0.3--12 keV) band due to the high $z$ of the source.
Furthermore the paucity of information about the characteristics of
\rbs~at other wavelengths in the literature does not allow to put any
additional  useful constraints on the nature of the absorber.
However, the X-ray constraint on the ionization status 
matches well with those reported so far for other warm absorbers
possibly detected in
high-$z$ Blazars (e.g. W04; F03; Worsley et al. 2004b) where the
obscuring matter was found to be nearly neutral or with loosely constrained $\xi$ \simlt~50
\cgs.
The inferred column density of the absorbing material is  \nh~$\approx$ a few
$\times$ 10$^{22}$ \cm2. Remarkably, such a value is very similar to
those reported from  \xmm~observations of all the other obscured
Blazars at high $z$ for which,
bearing in mind the small number statistics, the 
measured column densities of the absorbers seem 
to narrowly cluster around \nh~ $\approx$ 10$^{22}$~cm$^{-2}$
(e.g. F03; W04; Yuan et al. 2005; Grupe
et al 2005).
\nh~values \simlt~10$^{22}$ \cm2~are, however, hard to detect at high
redshifts and, hence, a selection effect is likely present.

Soft-energy cut-offs are more common 
in radio-loud objects (Fiore et al. 1998).  
One may therefore speculate that a mechanism of
 collisional ionization might be at work by multiple shocks along the
 jet (e.g. Gupta et al. 2005).  A similar phenomenon is believed to be present in the NLR of
 nearby Seyfert galaxies which are spatially associated with their
 radio jet structure, suggesting that the NLR originates from the
 compression due to the interaction between the outflowing radio
 material and the ambient gas in the galaxy (Capetti et al. 1996).
As the Blazar X-ray emission is dominated by relativistically beamed
 components from the jet, it is very likely that the obscuration may
 be due to jet--linked material and physical processes.\\

Our result strengthens the evidence for the existence
of a population of superluminous Blazar-like AGNs at $z$ $>$
2 which show soft X-ray cutoffs due to the likely presence of large amounts
of intrinsic absorbing matter with similar properties.  The discovery
of these jet-dominated  superluminous (i.e. L$_{\rm 0.5-10}$ \simgt~
10$^{47}$ \ergs) obscured QSOs at high $z$ is also important in the
context of the formation and cosmological evolution of radio-loud
objects.  Soft X-ray spectral cutoffs seems in fact to be a
prerogative of radio-loud QSOs (Fiore et al. 1998): {\it could the
gaseous environments in powerful Blazars be different from that in
radio-quiet AGNs?  Is the absorber intimately linked with the presence
of relativistic jets?}

Important clues about the physical state and geometry of the absorber
would be given by more sensitive ultra-soft X-ray observations, for
instance carried out by the under-study ESA {\it XEUS} satellite
(e.g. Bleeker \& Mendez 2002).   Furthermore, more robust constraints
on the ionization state of the obscuring matter may be also  achieved
with optical and UV observations by the detection of spectral
features typical of an ionized absorber.
\begin{acknowledgements}
We thank the anonymous referee for a valuable report.
We are grateful to the staff members of the \xmm~Science Operations Center for their support.
EP thanks P.~Rodriguez and J. A.~Carter for helpful discussions.
\end{acknowledgements}

\end{document}